\documentclass[twocolumn,prl,showpacs,
amsmath,amssymb]{revtex4}
\usepackage{hyperref}
\usepackage{graphicx}
\usepackage{dcolumn}
\usepackage{bm}

\begin{document}
\title{Hydrodynamics within the Electric Double Layer on slipping surfaces}

\author{Laurent Joly$^{1}$, Christophe Ybert$^{1}$, Emmanuel Trizac$^{2}$, Lyd\'eric Bocquet$^{1}$
}
\email{lbocquet@lpmcn.univ-lyon1.fr}
\affiliation{
$^{1}$ Laboratoire P.M.C.N., UMR CNRS 5586, Universit\'e Lyon I, 69622 Villeurbanne, France\\
$^{2}$ L.P.T.M.S, UMR CNRS 8626, B\^atiment 100, Universit\'e Paris XI, 91405 Orsay, France}

\date{\today}

\begin{abstract}
We show, using extensive Molecular Dynamics simulations, that the dynamics
of the electric double layer (EDL) is very much dependent on the wettability of the
charged surface on which the EDL develops. For a wetting surface, 
the dynamics, characterized by the so-called Zeta potential, is 
mainly controlled by the 
electric properties of the surface, and our work 
provides a clear interpretation for 
the traditionally introduced immobile Stern layer. In contrast, the immobile layer
disappears for non-wetting surfaces and the 
Zeta potential deduced from electrokinetic effects is considerably amplified
by the existence of a slippage at the solid substrate.
\end{abstract}

\maketitle

\narrowtext

The electric double layer (EDL) is a central concept in the understanding of the
static and dynamical properties of charged colloidal systems. This notion was introduced in the
early works of Gouy, Debye and H\"uckel \cite{Lyk} to describe the
distribution of micro-ions close to a charged colloidal surface. 
The EDL width determines the
electric interaction range between macromolecules and therefore controls
the static phase behaviour of these systems.  
On the other hand, at the dynamical level the EDL is at the origin
of numerous electrokinetic effects \cite{Lyk} : electrophoresis, electro-osmosis, streaming 
current or potential, etc. Because these various phenomena 
are governed by the {\it surface} of the sample
via the EDL, they provide smart and particularly efficient ways to 
drive or manipulate flows in micro-fluidic devices \cite{Stone,Bazant}, 
where surface effects become predominant. 

The extension of the EDL is typically on the order of a few nanometers and electrokinetic
phenomena therefore probe the {\it nanorheology} of the solvent+ions system at the
charged surface. This raises therefore some doubts about the validity of continuum 
approaches to describe the dynamics at such scales. These doubts are particularly relevant 
concerning the traditional description of the EDL dynamics, which relies both on the mean-field 
Poisson-Boltzmann theory of the micro-ion clouds, but also on continuum 
hydrodynamics for the flow fields \cite{Lyk}. These two aspects are embodied
in the so-called {\it Zeta potential}, denoted $\zeta$, 
which is {\it traditionally} defined as the electric potential $V(z_s)$
computed at the surface of shear $z_s$, where the fluid velocity
{\it vanishes} 
\cite{Lyk,Netz,Qiao,Churaev}.
This definition of $\zeta$ however relies on the somewhat uncontrolled assumption of no-slip
boundary condition of the solvent at the solid surface \cite{Duf}.

This assumption has been critically revisited in the last years. Indeed, 
a lot of progress has been made recently in the understanding of
the rheology of fluids at small scales, thanks in particular to computer simulations,
such as Molecular Dynamics (see eg \cite{BB} and references therein), but mainly to 
the development of new experimental techniques, such as Surface Force Apparatus
(SFA) or Atomic Force Microscope (AFM) \cite{AFMSFA,Vino}. The
conclusions emerging from these studies are that, while continuum hydrodynamics 
are found surprisingly to remain valid {\it up to very small length scales}, the {\it no-slip}
boundary condition (BC) for the fluid velocity at the solid surface may be violated 
in many situations (see \cite{Vino,BB,AFMSFA,Duf} and refs. therein). Moreover, this violation of the usual
no-slip BC is found to be 
controlled by the wetting properties of the fluid on the solid surface : 
while the no-slip BC is fulfilled on hydrophilic surfaces, 
a finite velocity slip is measured on hydrophobic surfaces \cite{BB,Vino}.

In this Letter we show using Molecular Dynamics (MD) simulations that a finite slip effect 
for the solvent at a charged surface considerably enhances the measured electrokinetic effects. 
This results in an enhanced $\zeta$ potential, the origin of which 
lies in the dynamics of the solvent at the surface: 
the flow-induced current, driven by the
convection of micro-ions in the EDL, is much larger in the presence of velocity slip at the surface 
than for the corresponding situation with a no-slip BC. 
Moreover, while an immobile Stern layer develops close to the charged
surface on hydrophilic samples, and does not contribute to electrokinetic effects, this layer is 
absent on hydrophobic surfaces and does contribute in a significant 
way to charge transport.

We first precise our microscopic model and some details of the simulation procedure. 
The fluid system  (solvent and micro-ions) is confined between two parallel solid substrates,
composed of individual atoms fixed on a fcc lattice. The solvent and solid substrate
particles interact via Lennard-Jones (LJ) potentials,
\begin{equation}
v_{ij}(r)=4 \epsilon \left[ \left({\sigma \over r}\right)^{12} - c_{ij}  \left({\sigma \over r}\right)^{6} \right] 
\label{LJ}
\end{equation}
with identical interaction energies $\epsilon$ and molecular diameters $\sigma$. The
tuning parameters $c_{ij}$ allow to adjust the wetting properties of the fluid on the substrate
\cite{BB}~: for a given fluid-fluid cohesion $c_{FF}$, the substrate exhibits a 
``hydrophilic'' behavior for large fluid-solid cohesivity, $c_{FS}$, and a ``hydrophobic'' behavior for small $c_{FS}$.
Here, the wetting (respectively non-wetting) situation is typically achieved by taking $c_{FS}=1$ (resp. 0.5)  for a fixed $c_{FF}=1.2$. This leads to a contact angle $\theta$ of a liquid droplet on the substrate, measured in the simulations equal to $80^\circ$ (resp. $140^\circ$) for a temperature $k_B T/\epsilon=1$ (we refer to \cite{BB} for an exhaustive discussion on this point). 
On the other hand, micro-ions interact both through LJ potentials,
as described in Eq. (\ref{LJ}), and Coulomb potential in a medium with
dielectric permittivity $\epsilon_d$
[$v_{\alpha\beta}(r)=k_B T q_\alpha q_\beta\ell_B/r$
where $q_\alpha$, $q_\beta$ are the valences of the interacting charges, 
and  $\ell_B= e^2/(4\pi \epsilon_d k_BT)$ is the 
Bjerrum length, $e$ denoting the elementary charge]. 
In water at room temperature $\ell_B= 0.7 nm$.  We shall choose in our simulations
$\ell_B=\sigma$.
Ions also interact directly
with the surface atoms via the same LJ potential as the
solvent ({\it i.e.} same $c_{FS}$). 
Micro-ions and solvent particles
have the same size $\sigma$ (in Eq. (\ref{LJ})).
Wall atoms
are organized into five layers of a fcc solid (100 direction) in both walls.
For each wall, the first layer only, in contact with the fluid, is charged.
The corresponding $N_w$ atoms bare a discrete charge, with valency
$q_{wall}=-Z/N_w$ so that each wall bears a negative net charge $-Z e$.
The solvent contains $Z$ monovalent counter-ions, 
to which $N_s=N_++N_-$ salt ions are added, all with valence one.
Global electroneutrality is enforced by imposing $N_+=N_-$. 
The systems simulated  are generally made up of $10^4$ atoms.
A typical solvent density is $\rho_f \sigma^3\sim 1$, while the concentration of
micro-ions $\rho_s=N_\pm/V$ is varied between $\rho_s \sigma^3= 5.10^{-3}$ and $\rho_s \sigma^3= 0.16$ (with $V$ the total volume of the sample).
This corresponds typically to an ionic strength varying between $10^{-2} M$ and $1 M$.
The corresponding Debye screening length (see below) is accordingly of the order of a few
Bjerrum lengths.
The charge per unit surface on the wall is $\Sigma=-0.2 e.\sigma^{-2}$, 
associated to
a Gouy-Chapman length $\lambda = (2\pi l_\mathrm{B} \Sigma)^{-1} = 0.8 \sigma$.
Using a typical value $\sigma=5\,$\AA, this corresponds roughly to $-0.13 C.m^{-2}$. 
Periodic boundary conditions are applied in the $x$ and $y$ directions, 
with $L_x=L_y=16 \sigma$ while the distance between the walls is $L_z= 20.8 \sigma$. 
Ewald sums are used to compute Coulombic interactions (assuming a periodicity in
the $z$ direction with a box size of $112 \sigma$, much larger than the wall to wall distance)
\cite{Plimpton}. Lennard-Jones units are used in the following
[distance $\sigma$, time $\tau=(m\sigma^2/\epsilon)^{1/2}$]. Temperature was kept
constant by applying a Hoover thermostat 
to the $y$ degrees of freedom, {\it i.e.} in the direction perpendicular to the flow and confinement.

Our model therefore includes the discrete nature of the solvent and charges and a tuning wettability of the surface,
whereas these effects are usually neglected in 
the traditional description of electrokinetic phenomena. Note that we chose 
to describe the charge interaction at the level of an effective dielectric media (with dielectric permittivity $\epsilon_d$). This 
simplifying assumption --that could be relaxed using a more realistic 
model for the solvent \cite{Qiao}--
allows to investigate specifically the generic interplay 
between slip effects and electric behavior, which is the main focus of this
work. We have moreover explored situations beyond the ones presented here 
(with or without salt, 
using different $c_{FS}$ for the ions and the solvent, and various Bjerrum
lengths $\ell_B$),
leaving the present conclusions unaffected.

\begin{figure}[t]
\begin{center}
\includegraphics[width=6cm,height=5cm]{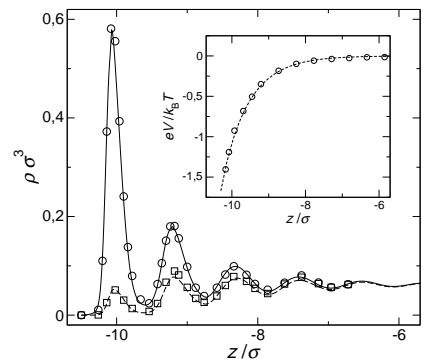}
\caption{Micro-ionic density profiles, averaged over the $xy$ directions 
($\rho_s\sigma^3=0.07$, wetting case). Symbols : MD simulations results 
for the counter-ions ($\circ$) and co-ions ($\scriptscriptstyle\square$); 
Solid and dashed lines : corresponding predictions of the 
modified PB description (see text).
Inset: Electrostatic potential. The wall is located at $z_w= -10.8 \sigma$. 
Symbols ($\circ$): MD simulation results calculated from Poisson's equation and micro-ions profiles; Dashed line~: bare 
PB prediction (see text).
}
\label{fig1}
\end{center} 
\end{figure}

We now turn to the simulation results. We first focus on the equilibrium properties of the
EDL. In Fig. \ref{fig1}, we show typical density profiles. 
The latter are found to exhibit important structuration effects close to the charged surface
and thus depart strongly from the Poisson-Boltzmann (PB) prediction \cite{Lyk}.
However, the oscillations in the micro-ions profiles originate in the structuration in the solvent itself, and
such an effect can be captured by a modified PB description. Indeed, due to the presence
of the solvent, micro-ions not only organize due to electric interactions
(which corresponds to the usual PB description) but also due to the effective external field 
associated with the structuration in the solvent \cite{Marcelja}, defined as $V_{ext}(z)=-k_B T \log\left[\rho_f(z)/\rho_f\right]$,
with $\rho_f(z)$ the solvent density profile and $\rho_f$ its bulk value. 
The micro-ions density profiles $\rho_\pm(z)$ correspondingly 
obey a modified Boltzmann equilibrium~:
\begin{equation}
\rho_\pm(z) \propto e^{\beta(\mp  e V(z) - V_{ext}(z)) }\propto \rho_f(z) e^{\mp \beta e V(z)}
\label{rhopm}
\end{equation}
with $\beta=1/k_BT$ and $V(z)$ the electrostatic potential.
Actually, such a relationship emerges naturally from a 
simple Density Functional Theory, in which the discrete nature 
of both solvent and charged atoms is taken into account exactly, 
while the standard mean-field
PB free energy is assumed for the electrostatic part. 
Using Poisson's equation, the electrostatic potential is found to obey a modified PB equation,
$\beta e \Delta V = \kappa^2 \gamma(z) \sinh(\beta e V)$
where $\kappa^2=8 \pi \ell_B \rho_s^{bulk}$ is the Debye screening factor defined in terms of the bulk
micro-ion concentration, and $\gamma(z)=\rho_f(z)/\rho_f$ is the normalized {\it solvent} density
profile. 
In order to test the predictions of this approach, we have measured the
fluid density profiles, $\rho_f(z)$,
and solved 
Poisson's equation with the microionic densities given by Eq. (\ref{rhopm}),
using Neumann boundary conditions, and assuming a smeared (uniform) surface 
charge on the wall.
As shown in Fig. \ref{fig1}, this
approach leads to results in very good agreement with the simulations 
profiles. Moreover,
a further approximation can be proposed: The solution of the modified PB equation 
for electrostatic potential is
actually very well approximated by the ``bare'' PB solution $V_{PB}(z)$ (corresponding
to $\gamma(z)=1$), whose analytic expression can be found in the literature \cite{Lyk}.
This leads to $\rho_\pm(z) \propto \rho_f(z) exp[\mp \beta e V_{PB}(z)]$.
The validity of this approximation --surprising in view of the strong layering 
effect at work--
is emphasized in Fig. \ref{fig1} (inset), where the corresponding 
bare PB potential \cite{Lyk} is plotted 
against the
"exact" electric potential. The latter is  
obtained from the simulations using Poisson's equation by 
integrating twice the charge density profile $\rho_C=e(\rho_+-\rho_-)$.

We now investigate the dynamical properties of the EDL. We first consider a {\it streaming current}
experiment: an external volume force, $f_0$, is applied to the fluid in the $x$ direction, enforcing a 
Poiseuille flow in the cell, and 
the electric current, $I_e$, associated with the convective motion of the micro-ions is measured. 
The standard EDL description of this electrokinetic effect predicts a linear relationship between the
current and the force, in the form \cite{Lyk}~:
\begin{equation}
I_e=-{\epsilon_d \zeta \over \eta} {\cal A} f_0
\label{Ie}
\end{equation}
where $\eta$ is the shear viscosity of the fluid and
${\cal A}$ the fluid slab cross area. In the simulations, a force per particle is applied to all fluid particles
and the corresponding electric current is measured. We emphasize that
linear response (in the applied force) of the system was carefully checked. 
{\it In the following we shall use this expression as the definition of the - apparent - $\zeta$ potential}, in 
line with experimental procedures.

\begin{figure}[b]
\includegraphics[width=6cm,height=5cm]{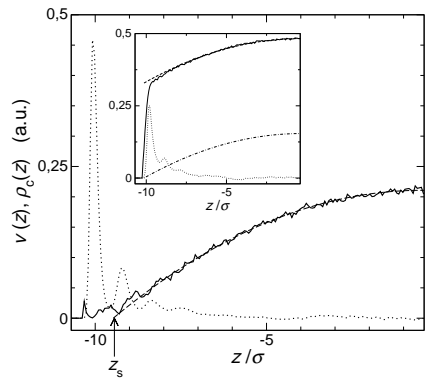}
\caption{
Measured Poiseuille velocity profile (solid line) in the wetting case ($c_{FS}=1$). Dashed line : hydrodynamic prediction
using a no-slip BC at the 'plane of shear' located at $z_s$ (indicated by the arrow). 
To emphasize the existence of 
an immobile Stern layer, we also indicate the charge density
profile $\rho_C(z)=e(\rho_+(z)-\rho_-(z))$ (dotted line), with arbitrary units.
The position of the wall (defined as that of
the centers of the last layer of wall atoms) 
is at $z_w=-10.8 \sigma$.
Inset: Results for the non-wetting case ($c_{FS}=0.5$). Solid line~: velocity profile measured in the simulation (shown on the same scale as
in the main graph); Dashed line~: hydrodynamic prediction with a partial slip BC, with a slip length
$b=11 \sigma$; Dashed-dot line~: hydrodynamic prediction with a no-slip BC; Dotted line~: 
charge density profile (arbitrary units).
}
\label{fig2}
\end{figure}

We first discuss the measured velocity profiles. 
The situation corresponding to a wetting substrate 
is shown in the main plot of Fig. \ref{fig2} (here for $f_0=0.02$ in LJ units).
The velocity profile is found to exhibit a parabolic shape as predicted by continuum
hydrodynamics, even at the scale of the EDL. Moreover the viscosity, deduced from the curvature of 
the parabolic shape, is measured to keep its bulk value. Nevertheless, the no-slip BC is found to
apply inside the liquid, {\it at a distance of about one layer of solvent particle}, in agreement
with previous theoretical predictions \cite{BB}. This position of the
no-slip BC here defines the ``plane of shear'' position, $z_s$, usually introduced in the
electrokinetic literature \cite{Lyk}. 
As shown in Fig. \ref{fig2}, 
the layer of micro-ions located within $z_s$, does not contribute to
the convective transport, thereby reducing the global streaming current. This
immobile layer coincides with the so-called Stern layer of immobile micro-ions close
to the charged surface \cite{Lyk}. Note moreover that we found $z_s$ to vary slightly with
electrostatic parameters (surface charge $\Sigma$, Bjerrum length $\ell_B$), as expected.

On the other hand, the non-wetting case 
exhibits a very different behaviour, as shown in the
inset of Fig. \ref{fig2}. 
First, concerning the velocity profile, a large amount of slip is
found at the wall surface, in agreement with observations on uncharged non-wetting surfaces \cite{BB}.
More quantitatively,
slippage is characterized by a {\it slip length}, $b$, defined as the distance at
which the linear extrapolation of the velocity profile vanishes. 
This amounts
to replacing the no-slip BC by a partial slip BC, defined as $b {\partial v\over \partial z} = v$ 
at the wall position \cite{BB}. As shown in the inset of Fig. \ref{fig2}, the velocity profile is found to
be well fitted by the continuum hydrodynamics (parabolic) prediction, together with a partial slip 
BC, characterized by a non vanishing slip length (here $b \simeq11 \sigma$). 
An important point here is that the first layer of micro-ions now
contributes a large amount to the global streaming current in contrast to the wetting case,
as emphasized in the inset of Fig. \ref{fig2}. 
The remobilization of the Stern layer therefore adds on to the slippage effect to increase the
$\zeta$ potential.

\begin{figure}[htb]
\includegraphics[width=6cm]{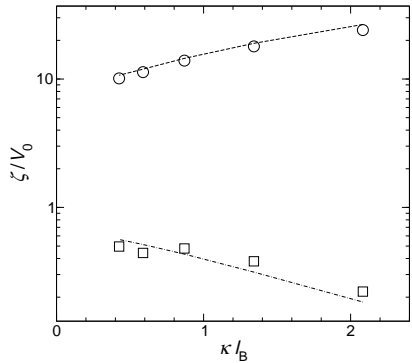}
\caption{The symbols show the $\zeta$ potential measured in MD as a function the screening factor $\kappa\ell_B$,
for the wetting (bottom) and non-wetting substrate (top). The $\zeta$
potential is normalized by the bare surface potential $V_0$ obtained from the PB 
expression at a given $\kappa$ and surface charge (see the discussion
on the inset of Fig. \ref{fig1}).
For the wetting case (bottom), the dashed line is the PB 
electrostatic potential $V(z_s)$ where
the 'plane of shear' position $z_s$ does not vary significantly with salt. 
For the non-wetting case (top), the dash-dotted
line corresponds to the slip prediction, 
Eq. (\ref{zetaeff}) (with $b=11 \sigma$). 
}
\label{fig3}
\end{figure}

We now summarize our results in Fig. \ref{fig3} and plot the $\zeta$ potential
[deduced from the measure of the charge current supplemented with Eq. (\ref{Ie})]
as a function of the Debye screening factor in the wetting and non-wetting cases.
In this
figure the $\zeta$ potential is normalized by the bare surface potential $V_0$,
obtained from the analytic PB expression
\cite{Lyk} (see Fig. \ref{fig1}). 
Moreover, the slip length
in the non-wetting case has been measured to barely depend on the screening
factor, indicating that the micro-ions do not affect the 
fluid-solid friction. 
The overall conclusion from Fig. \ref{fig3} is that non-wettability strongly amplifies
the electrokinetic effects : the ratio between the $\zeta$ potential and the surface potential
is much larger in the hydrophobic case as compared to the hydrophilic case. 
More precisely, in the wetting case
the $\zeta$ potential is fixed by the electric properties of the surface, 
and coincides with the electric potential at the 'plane of shear', $\zeta \simeq V(z_s)$,
as is usually assumed \cite{Lyk}. This is demonstrated in Fig. \ref{fig3}, where 
the simulation points for $\zeta$ are compared to the PB estimate for the
electric potential, $V_{PB}(z_s)$, showing an overall very good agreement. 
Conversely, the $\zeta$ potential in the non-wetting case is dominated by the slip effect
and the immobile Stern layer is completely absent. The effect of slip can be accounted
for by considering the partial slip BC in the electrokinetic 
current $I_e=\int dS \rho_C(z) v(z)$ with $\rho_C(z)$ the charge density
and $v(z)$ the velocity profile characterized by a slip length $b$. Within linearized
PB description, and for the present planar geometry, the result for the
current $I_e$, Eq. (\ref{Ie}), may then be written  
$I_e={\epsilon_d V_0\over \eta} (1 + \kappa b) f_0$ \cite{Stone}.
In the non-wetting case, this leads to
\begin{equation}
\zeta= V_0 (1 + \kappa b)
\label{zetaeff}
\end{equation}
with $V_0$ the bare potential of the surface \cite{note}. This expression 
is successfully compared in
Fig. \ref{fig3} to simulation results.

Finally we quote that we have tested a different electrokinetic geometry, corresponding
to the more common electro-osmotic situation~: flow response to the application of an
electric field. The $\zeta$ potentials measured in this geometry (not shown here) are in full
{qualitative} and {quantitative} agreement with the results in the present (streaming current) 
geometry, both on the wetting and non-wetting surfaces. 

To conclude, we have shown using MD simulations 
that the notion of $\zeta$ potential, the cornerstone in the description of 
EDL dynamics, encompasses different physical mechanisms, depending on the
wettability of the charged substrate. 
In the wetting situation, the $\zeta$ potential can be indeed directly related to surface charge properties, confirming hereby the traditional Stern layer picture \cite{Lyk}. 
In contrast, for non-wetting substrates, electric and slip effects are strongly intricated, 
leading for the $\zeta$ potential (Eq. \ref{zetaeff}) to an amplification ratio of $ 1+ \kappa b$. Practically, this ratio can take large
values, even for moderate slip lengths on the order of a few nanometers, 
since the Debye length 
$\kappa^{-1}$ usually lies in the nanometer range
(to be specific, slip lengths of the order of tens of nanometers have been reported for hydrophobic silanized glass surfaces \cite{Vino}). It would be moreover highly desirable to 
generalize Eq. (\ref{zetaeff}) obtained for the planar case to
other geometries (eg. spherical) where a new length (radius $R$) will compete with the debye
($\kappa^{-1}$) and slip ($b$) lengths, Eq. (\ref{zetaeff}) yielding the limiting behavior for $b/R, 
\kappa^{-1}/R \rightarrow 0$.
This work also points to the difference between dynamical and static properties of the EDL,
respectively characterized by the Zeta and bare surface potential, $\zeta$ and $V_0$. 
Independent static and dynamic measurements, using {\it e.g.} AFM or SFA, should therefore allow 
to probe the different mechanisms underlined in this study. Another route, with potential applications 
in micro-fluidics, consists in polarizing the solid substrate, {\it i.e.} ``imposing'' $V_0$, which
should lead to large electrokinetic effects on non-wetting substrates.




\end{document}